\documentclass[twoside,twocolumn,english]{revtex4-1}
\usepackage[plainpages=false,colorlinks=true,linkcolor=blue, citecolor=blue, urlcolor=blue]{hyperref}
\usepackage[english]{babel}
\usepackage[T1]{fontenc}
\usepackage[utf8]{inputenc}
\usepackage{amsmath}
\usepackage{graphicx}
\usepackage[utopia]{mathdesign}
\usepackage{bm}

\makeatletter

\newcommand\dgf{{\phantom{\dagger}}}

\newcommand\Rv{\mathbf{R}}
\newcommand\rv{\mathbf{r}}
\newcommand\Qv{\mathbf{Q}}

\newcommand\Sigmav{\mathbf{\Sigma}}
\newcommand\Gv{\mathbf{G}}
\newcommand\kvt{\mathbf{\tilde k}}
\newcommand\xh{\mathbf{x}}
\newcommand\yh{\mathbf{y}}
\newcommand\Deltab{\skew1\bar\Delta}
\newcommand\fb{\skew3\bar f}

\@ifundefined{date}{}{\date{\today}}
\makeatother

\begin{document}
\title{CDMFT+HFD : an extension of dynamical mean field theory for nonlocal interactions applied to the single band extended Hubbard model}
\author{Sarbajaya Kundu$^{1}$ and David S\'en\'echal$^{2}$}
\address{$^{1}$Department of Physics, University of Florida, Gainesville, FL 32611, USA}
\address{$^{2}$D\'epartement de physique and Institut quantique, Universit\'e de Sherbrooke, Sherbrooke, Qu\'ebec, Canada J1K 2R1}

\begin{abstract}
We examine the phase diagram of the extended Hubbard model on a square lattice, for both attractive and repulsive nearest-neighbor interactions, using CDMFT+HFD, a combination of Cluster Dynamical Mean Field theory (CDMFT) and a Hartree-Fock mean-field decoupling of the inter-cluster extended interaction. 
For attractive non-local interactions, this model exhibits a region of phase separation near half-filling, in the vicinity of which we find pockets of $d$-wave superconductivity, decaying rapidly as a function of doping, with disconnected patches of extended $s$-wave order at smaller (higher) electron densities. 
On the other hand, when the extended interaction is repulsive, a Mott insulating state at half-filling is destabilized by hole doping, in the strong-coupling limit, in favor of $d$-wave superconductivity.
At the particle-hole invariant chemical potential, we find a first-order phase transition from antiferromagnetism (AF) to $d$-wave superconductivity as a function of the attractive nearest-neighbor interaction, along with a deviation of the density from the half-filled limit. 
A repulsive extended interaction instead favors charge-density wave (CDW) order at half-filling. 
\end{abstract}
\maketitle

\section{Introduction}

The single-band Hubbard model has long served as a useful platform for studying the effect of strong electronic correlations
\citep{kao_unified_2022,li_short_2021,qin_hubbard_2022,tasaki_hubbard_1998,arovas_2022,Izyumov_1995}.
In particular, it explains many of the experimental observations in the high-$T_{c}$ cuprate superconductors~\citep{armitage_progress_2010,atikur_rahman_review_2015,chu_hole-doped_2015,fradkin_colloquium_2015,li_short_2021,maksimov_high-temperature_2000,norman_electronic_2003,orenstein_advances_2000,ruvalds_theoretical_1996,shen_cuprate_2008,varma_colloquium_2020}, providing an approximate picture for the description of these materials~\citep{aichhorn_superconducting_2007,bulut_quasiparticle_1994,wermbter_self-consistent_1993,simon_optical_1997,simkovic_origin_2022,sheshadri_connecting_2023,macridin_physics_2005,kuroki_link_1999,mushoff_magnetic_2021}.
More recently, there have been numerous studies on extensions of this model with nearest-neighbor interactions, known as the extended Hubbard model (EHM)~\citep{aichhorn_charge_2004,ayral_screening_2013,calandra_metal-insulator_2002,callaway_small-cluster_1990,carvalho_formation_2022,chattopadhyay_phase_1997,chen_superconducting_2022,coppersmith_superconducting_1990,davoudi_nearest-neighbor_2006,davoudi_non-perturbative_2007,fresard_charge_2020,gilmutdinov_interplay_2022,hassan_slave_2010,huang_unconventional_2013,jedrzejewski_phase_1994,kagan_triplet_2011,kapcia_doping-driven_2017,kapcia_effects_2011,linner_coexistence_2023,linner_multi-channel_2022,littlewood_collective_1990,littlewood_pairing_1989,merino_nonlocal_2007,merino_superconductivity_2001,micnas_extended_1988,micnas_superconductivity_1988,micnas_superconductivity_1989,micnas_superfluid_2002,murakami_possible_2000,ohta_exact-diagonalization_1994,onari_phase_2004,peng_enhanced_2023,philoxene_spin_2022,pietig_reentrant_1999,plakida_theory_2013,plonka_fidelity_2015,pudleiner_competition_2019,raghu_effects_2012,roig_revisiting_2022,rosciszewski_2003,rosciszewski_pair_1995,schuler_first-order_2018,schuler_thermodynamics_2019,sherman_two-dimensional_2023,sousa-junior_phase_2023,su_phase_2004,su_spin-density_2001,sun_linear_2023,sushchyev_thermodynamics_2022,szabo_superconducting_1996,terletska_charge_2017,terletska_charge_2018,terletska_dynamical_2021,tong_charge_2004,van_dongen_thermodynamics_1991,van_loon_extended_2018,vandelli_dual_2020,vojta_phase_2001,wang_spectral_2022,wolf_unconventional_2018,yan_theory_1993,yao_determinant_2022,yoshimi_enhanced_2009,zhang_extended_1989,zhou_robust_2023}.
There are several reasons for the continuing interest of the community in exploring the effect of non-local interactions. 
In actual materials, the interactions between neighboring sites may not be completely screened,
necessitating a more careful treatment of longer-range interactions.
The model with an attractive nearest-neighbor interaction provides an effective representation of the attractive interactions mediated by electron-phonon coupling, and may be realized in ultra-cold atom systems.
The relevance of studying such a model is further emphasized by recent ARPES studies on the one-dimensional cuprate chain compound Ba$_{2-x}$Sr$_{x}$CuO$_{3+\delta}$~\citep{chen_anomalously_2021}, where the observations can be explained using a Hubbard model with an attractive extended interaction.
On the other hand, the model with repulsive non-local interactions provides an ideal playground for studying the interplay of charge and spin fluctuations, since the relative magnitude of the charge fluctuations can be controlled by the strength of the extended interaction~\citep{aichhorn_charge_2004,carvalho_formation_2022,davoudi_nearest-neighbor_2006,davoudi_non-perturbative_2007}.
The EHM at quarter-filling has proven useful for describing the charge ordering transition due to inter-site Coulomb interactions in a variety of materials
~\citep{calandra_metal-insulator_2002,vojta_phase_2001,merino_nonlocal_2007,merino_superconductivity_2001,tong_charge_2004}.
Both the Hubbard model and its extension with longer-range interactions have contributed significantly to the methodological development in the field of strongly correlated systems, and in particular high-$T_{c}$ superconductors, which is essential for obtaining results that can be quantitatively compared with experiments.

In recent years, the properties of the EHM have been analyzed using a variety of approaches, including, among others, mean-field theory~\citep{micnas_extended_1988,micnas_superconductivity_1988,micnas_superconductivity_1989,su_spin-density_2001}, 
functional renormalization group (fRG)~\citep{huang_unconventional_2013}, exact diagonalization (ED)~\citep{chen_superconducting_2022,plonka_fidelity_2015,callaway_small-cluster_1990,ohta_exact-diagonalization_1994}, density-matrix renormalization group (DMRG)~\citep{peng_enhanced_2023,raghu_effects_2012}, Quantum Monte Carlo (QMC)~\citep{yao_determinant_2022,sousa-junior_phase_2023,zhang_extended_1989,jiang_d_2018}
and the fluctuation-exchange approximation (FLEX)~\citep{onari_phase_2004}.
However, many of the approaches used are best suited for studying the weak-coupling or the strong-coupling limit, and there are few that can describe the intermediate-coupling regime equally well.
Even among those that can, each has it own limitations. For instance, simple exact diagonalizations are restricted to small systems, quantum Monte Carlo methods suffer from the fermion sign problem in many applications of interest, the density-matrix renormalization group (DMRG) applies to one-dimensional or ribbon-like systems, etc.
In addition, certain aspects of the model with repulsive interactions have been studied in detail using the so-called extended dynamical mean-field theory (EDMFT) approach~\citep{hu_extended_2022,huang_extended_2014,chitra_effect_2000}, 
in which the local density fluctuations together with the local self-energy are propagated on the whole lattice using the known dispersion and density-density extended interactions.
Other variations of this method, such as a combination of EDMFT with the GW approximation~\citep{ayral_screening_2013,ayral_spectral_2012,sun_extended_2002,sun_many-body_2004},which perturbatively includes non-local self-energy corrections, and the dual boson method~\citep{vandelli_dual_2020,van_loon_beyond_2014,van_loon_extended_2018}, which constructs a diagrammatic expansion about the extended DMFT, have likewise contributed to its understanding.
More recently, cluster methods~\citep{aichhorn_charge_2004,hassan_slave_2010,kotliar_cellular_2001,terletska_charge_2017,terletska_charge_2018,terletska_dynamical_2021,tong_extended_2005}, which capture short-range correlations non-perturbatively within periodic clusters, have also been applied to this model.
However, such studies have largely been limited to fixed densities and repulsive interactions.
Overall, there have been fewer studies that consider both an extensive range of interaction couplings and band fillings, and relatively less focus on the case of attractive extended interactions.

In this paper, we study the phase diagram of the extended Hubbard model on a square lattice, for both attractive and repulsive nearest-neighbor interactions, using CDMFT+HFD, an extension of the Cluster Dynamical Mean Field Theory (CDMFT)~\citep{lichtenstein2000,kotliar_cellular_2001} approach with a 
Hartree-Fock decoupling of the inter-cluster interactions.
CDMFT belongs to a class of methods called Quantum Cluster Methods~\citep{maier2005,adebanjo_ubiquity_2022,avella_cluster_2012,avella_cluster_2012-1,senechal_variational_2008,senechal_introduction_2010,slezak_multi-scale_2009}, which are a set of closely related approaches that consider a finite cluster of sites embedded in an infinite lattice, and introduce additional fields or ``bath'' degrees of freedom, so as to best represent the effect of the surrounding infinite lattice.
The values of these additional parameters are determined by using variational or self-consistency principles.
These methods have proven useful for interpolation between results obtained in the weak- and strong-coupling regimes, since their accuracy is controlled by the size of the clusters used, rather than the strength of the couplings.
Further, we treat the inter-cluster interactions within a Hartree-Fock mean-field 
decoupling, which generates additional Hartree, Fock and anomalous contributions to the cluster Hamiltonian.
While a similar treatment has been used to study the model at quarter-filling~\citep{merino_nonlocal_2007} for the case of repulsive interactions, with the objective of understanding the electronic properties of metals close to a Coulomb-driven charge ordered insulator transition,
this analysis was focused on a specific parameter regime, and did not include superconducting orders. 

This work constitutes a test of the  CDMFT+HFD method, described in Sec.II below. Our main findings are as follows.
For a weak repulsive local interaction $U$ and an attractive extended interaction $V$, the system undergoes a transition towards a phase separated (PS) state when the chemical potential lies in the vicinity of its particle-hole symmetric value, $U/2+4V$.
The exact region of phase separation is identified by using the hysteresis in the behavior of the electron density as a function of the chemical potential, which corresponds to the coexistence of two different uniform-density solutions.
As a function of doping away from the half-filled point, symmetrical and sharply decaying regions of $d_{x^2-y^2}$-wave superconducting order are observed, followed by disconnected pockets of extended $s$-wave order near quarter-filling, as well as at very small (large) densities.
A stronger attractive extended interaction tends to favor phase separation as well as superconductivity, whereas the repulsive on-site interaction $U$ is found to be detrimental 
to both.
At the particle-hole symmetric chemical potential, we detect a first-order phase transition from antiferromagnetism (AF) to $d$-wave superconductivity as the attractive $V$ becomes stronger, which is accompanied by a gradual deviation of the density from its half-filled limit, induced by phase separation.
For repulsive nearest-neighbor interactions in the strong-coupling regime $U\gg t$, the Mott insulating state at half-filling is destabilized, upon hole doping, in favor of a dome-shaped region of $d$-wave superconducting order.
This order is found to be remarkably stable in the presence of a non-local interaction, and slightly suppressed by it.
At half-filling, a repulsive non-local interaction induces a first-order phase transition from antiferromagnetism (AF) to a charge-density wave (CDW) order.
Our results are qualitatively in agreement with the existing literature on the phase diagram of the EHM, with some notable differences in the region of attractive interactions.
An important difference is that intra-cluster fluctuations are treated exactly, which tends to make superconducting orders somewhat weaker in this approach.

The paper is organized as follows.
In Sect.~II, we introduce the model Hamiltonian, and provide a brief overview of the CDMFT approach that we use for our analysis, as well as the Hartree-Fock mean-field decoupling of the intercluster interactions.
In Sect.~III, we describe
the phase diagram obtained as a function of the interaction strength and doping, and the phase transitions observed at half-filling.
Finally, in Sect.~IV, we summarize our results, discuss some relevant observations and present the conclusions of our study.

\section{Model and method}

\subsection{Model Hamiltonian}

The general form of the extended Hubbard model Hamiltonian is 
\begin{multline}
H=\sum_{\rv,\rv',\sigma}t_{\rv\rv'}c_{\rv\sigma}^{\dagger}c_{\rv'\sigma}^\dgf + U\sum_{\rv}n_{\rv\uparrow}n_{\rv\downarrow}\\
+\frac{1}{2}\sum_{\rv,\rv',\sigma,\sigma'}V_{\rv\rv'}n_{\rv\sigma}n_{\rv'\sigma'}
\end{multline}
where $\rv,\rv'$ label lattice sites, $t_{\rv\rv'}$ are the hopping amplitudes, $U$ the on-site Hubbard interaction, and $V_{\rv\rv'}$ the nearest-neighbor interaction (each bond counted once, hence the factor $\frac12$).

For the purpose of our analysis, we study the following model on a square lattice:
\begin{multline}
H =-t\sum_{\rv}\left( c_\rv^{\dagger}c_{\rv+\xh}^\dgf + c_\rv^{\dagger}c^\dgf_{\rv+\yh} + \mathrm{H.c.}\right)
+ U\sum_{\rv}n_{\rv\uparrow}n_{\rv\downarrow}\\
 -\mu\sum_{\rv}(n_{\rv\uparrow}+n_{\rv\downarrow})+V\sum_{\rv,\sigma,\sigma'}\Big(n_{\rv\sigma}n_{\rv+\xh,\sigma'}+n_{\rv\sigma}n_{\rv+\yh,\sigma'}\Big)
\end{multline}
where $\xh,\yh$ are the lattice unit vectors along the $x$ and $y$ directions, and the operator $c_{\rv\alpha}$ annihilates a particle with spin $\alpha=\uparrow,\downarrow$ at site $\rv$. The occupation number is $n_{\rv\alpha}=c_{\rv\alpha}^{\dagger}c^\dgf_{\rv\alpha}$.
We consider a range of values for the chemical potential $\mu$, corresponding to a continuous range of densities, from $n=0$ to $2$, along with a repulsive local interaction $U>0$, and a nearest-neighbor interaction $V$ that can be positive or negative.
The particle-hole symmetric value of the chemical potential, $\mu=U/2+4V$, which corresponds to a half-filled band in the absence of phase separation, features prominently in our analysis.
The unit of energy is taken to be the nearest-neighbor hopping amplitude $t=1.0$, with the lattice constant $a=1$.
Note that in the absence of longer-range hopping terms, beyond the nearest-neighbor bonds, the model respects particle-hole symmetry $n\to 2-n$.

We examine the possibility of superconducting as well as density-wave orders.
For this purpose, the anomalous operators are defined on the lattice using a $d$-vector, as 
\begin{equation}
\Delta_{\rv\rv',b} c_{\rv s}(i\sigma_{b}\sigma_{2})_{ss'}c_{\rv's'}+\mathrm{H.c.}
\end{equation}
where $b=0,1,2,3$, and $\sigma_{b}$ are the Pauli matrices.
The case $b=0$ corresponds to singlet superconductivity, in which case $\Delta_{\rv\rv',0} = \Delta_{\rv'\rv,0}$ and the cases $b=1,2,3$ correspond to triplet superconductivity, in which case, $\Delta_{\rv\rv',b}=-\Delta_{\rv'\rv,b}$.
In practice, these operators are defined by specifying $b$ and the relative position $\rv-\rv'$. 

Density wave operators are defined with a spatial modulation characterized by a wave vector $\Qv$, and can be based on sites or on bonds. 
In our analysis, we focus on site density waves, defined as 
\begin{equation}
\sum_{\rv}A_{\rv}\cos(\Qv\cdot\rv+\phi)
\end{equation}
where $A_{\rv}=n_{\rv},S_{\rv}^{x},S_{\rv}^{z}$ corresponds to charge- or spin-density wave orders, and $\phi$ is a sliding phase.
We probe the presence of density-wave orders with $\Qv=(\pi,\pi)$ and $\phi=0$. 

\begin{figure}
\begin{centering}
\includegraphics[scale=1]{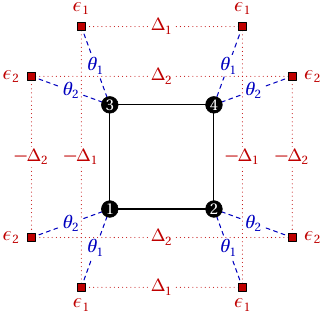}
\par\end{centering}
\caption{\label{fig:bath_simple}
Schematic representation of the first (``simple'') impurity problem used in our analysis, with bath energies $\epsilon_{i}$, cluster-bath hybridization parameters $\theta_{i}$ and anomalous bath parameters $\Delta_{i}$.
Physical sites are marked by numbered black dots and bath orbitals by red squares.
We choose the bath parameters such that the environment of each cluster site is identical.
This impurity model has reflection symmetry with respect to horizontal and vertical mirror planes ($C_{2v}$ symmetry), and typically involves only spin-independent hopping terms.
Pairing terms $\Delta_{1,2}$ are introduced between bath orbitals, with signs adapted to the SC order probed (shown here for a $d$-wave order, but all positive for an extended $s$-wave order). The number of independent bath parameters is 6.}
\end{figure}

\begin{figure}
\begin{centering}
\includegraphics[scale=1]{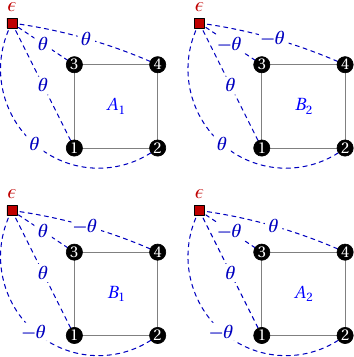}
\par\end{centering}
\caption{\label{fig:bath_general}
Schematic representation of the second (``general'') impurity problem used in our analysis.
Each representation of the point group $C_{2v}$ ($A_{1,2}$ and $B_{1,2}$) corresponds to a set of phases ($\pm 1$), and each of the 8 bath orbitals
belongs to one of these four representations (two bath orbitals per representation). The different bath orbitals are independent (the bath system is diagonal) and we only show here a view of each of the four representations with the corresponding signs associated to each cluster site (black dots).
The hybridization parameters $\theta$ are shown, and corresponding pairing operators (or anomalous hybridizations) between each bath orbital and each site also exist, with the same relative phases. The number of independent bath parameters is 18.}
\end{figure}

\subsection{Method: CDMFT+HFD}

Here, we provide a brief overview of the Quantum Cluster Method used in our analysis.
For a more detailed discussion of the basic principles and mathematical background of such methods, please see Ref.~\onlinecite{maier2005,senechal2023qcm,avella_cluster_2012}.

Cluster dynamical mean-field theory (CDMFT) is an extension of the dynamical mean-field theory (DMFT)\citep{georges_dynamical_1996,georges_strongly_2004,held_electronic_2007,vollhardt_dynamical_2020}
approach where, instead of a single-site impurity, we consider a cluster of sites with open boundary conditions, taking into account short-range spatial correlations exactly.
In this approach, a set of uncorrelated, additional ``bath'' orbitals hybridized with the cluster are used to account for the effect of the cluster's environment.
Thus, the infinite lattice is tiled into identical, small clusters and each of these is coupled to a bath of uncorrelated, auxiliary orbitals.
These bath orbitals have their own energy levels $\epsilon_{i\sigma}$, which may or may not be spin dependent, and are hybridized with the cluster sites (labeled $r$) with amplitudes $\theta_{ir\sigma}$ .
In addition, for studying superconducting orders, different types of anomalous pairings $\Delta_{ij\sigma\sigma'}$ may be introduced between bath orbitals $i,j$ or $\Delta_{ir\sigma\sigma'}$ between bath orbital $i$ and cluster site $r$.

In our analysis, we use two types of bath models. 
In the simple model (Fig.~\ref{fig:bath_simple}), the environment of each cluster is identical, and we introduce two bath orbitals per cluster site. 
Parameters of the impurity model include bath orbital energy levels ($\epsilon_{1,2}$), hybridization between each cluster site and the corresponding bath orbitals ($\theta_{1,2}$), and pairings between the bath orbitals ($\Delta_{1,2}$). 
The precise form of $\Delta_{1,2}$, including their relative phases between different bath orbitals, depends on whether we probe extended $s$-wave, $d$-wave, or triplet superconductivity. This simple impurity model involves 6 independent parameters to be determined self-consistently. At half-filling, we introduce bath energies as well as hoppings, that are consistent with the appearance of a density-wave order, and additionally spin-dependent in the presence of antiferromagnetism. This increases the number of independent parameters. However, imposing particle-hole symmetry at half-filling once again reduces this number to 6.  

We also use a more general bath model (Fig.~\ref{fig:bath_general}). While the total number of bath orbitals is unchanged, every bath orbital is connected to every cluster site (with distinct combinations of relative phases), and we define bath energies, cluster-bath hybridizations and anomalous pairings between the cluster and the bath sites. 
In this model the bath is diagonal, i.e., the different bath orbitals are not directly coupled between themselves, and, taking into account rotation symmetry, there are 18 independent bath parameters to set.
At at particle-hole symmetric point, we introduce bath energies, hybridizations and anomalous pairings that have two different values for alternative sites. Even upon taking into account particle-hole symmetry, this gives us a total of 52 independent parameters in the presence of superconductivity, and 20 independent parameters when superconductivity is absent (i.e. for $V>0$).

All bath parameters are determined by a self-consistency condition (see Ref.~\cite{maier2005, senechal2023qcm, avella_cluster_2012} for details). The simple bath model is expected to be easier to converge than the general bath model, because of the smaller set of parameters. While we expect the results obtained from the general bath model to be more reliable, we do find most of the results to be qualitatively similar in the two cases. 
Once the bath parameters are converged, the self-energy $\Sigmav(\omega)$ associated with the cluster is applied to the whole lattice, so that the lattice Green function is
\begin{equation}
\Gv^{-1}(\kvt,\omega)=\Gv_{0}^{-1}(\kvt,\omega)-\Sigmav(\omega)
\end{equation}
where $\kvt$ denotes a reduced wave vector (defined in the Brillouin zone of the super-lattice of clusters defined by the tiling) and $\Gv_{0}$ is the non-interacting Green function.
The Green-function-like objects $\Gv$, $\Gv_0$ and $\Sigmav$ are $L\times L$ matrices, $L$ being the number of physical degrees of freedom on the cluster (here $L=8$ because of spin and the four cluster sites).
The lattice Green function $\Gv$ determined using the solution for the optimum values of the bath parameters can be used for obtaining the average values of one-body operators defined on the lattice.
An exact-diagonalization solver is used at zero temperature (the Lanczos method or variants thereof), and the computational size of the problem is determined by the total number of cluster and bath orbitals, and increases exponentially with that number. 

In the presence of extended interactions, we also perform a Hartree-Fock mean-field decomposition of the interaction terms defined between different clusters, while the interactions within a cluster are treated exactly. The inter-cluster interactions are decoupled in the Hartree, Fock and anomalous channels, which contribute to the number density, the hopping and the pairing operators, respectively. Moreover, we only retain those combinations of the site/bond operators that are physically relevant in the regions we work in (such as $d-$wave or extended $s-$wave), and discard the rest. The mean-field values of the relevant combinations are determined self-consistently, within the CDMFT loop that optimizes the bath parameters.
For the details of this procedure, please refer to the Appendix.

\section{Results}

In this section, we discuss the salient features of the phase diagram obtained from our analysis, for both attractive and repulsive nearest-neighbor interactions.
The dominant superconducting and density-wave orders are identified by computing the corresponding order parameters 
using the optimum values of the CDMFT (bath and mean-field) parameters, as a function of electron density, as well as at half-filling.
In the following analysis, we focus our attention on the strong coupling limit $U\gg t$ for $V>0$, which is a regime well-understood on physical grounds.
For $V<0$, we consider relatively weak interactions $U\sim t$, far from the Mott insulating regime, which primarily serve the purpose of controlling the extent of phase separation when the interaction $V$ becomes sufficiently attractive.
At half-filling, we confirm the nature of the phase transitions, by plotting the relevant order parameters both as a function of $U>0$, for fixed values of $V>0$ or $V<0$, and as a function of $V$ for fixed values of $U$. 

\subsection{Phase diagram at the particle-hole symmetric chemical potential}

Here, we fix the chemical potential to $\mu=U/2+4V$, corresponding to a  half-filled band, and examine the behavior of different superconducting and density-wave orders, as a function of the local repulsion $U$ as well as attractive/repulsive $V$.
While antiferromagnetism is favored at half-filling, in both the weak- and strong-coupling regimes, an attractive non-local interaction is expected to drive the system towards a superconducting instability, and eventually phase separation.
On the other hand, repulsive interactions $V$ would typically foster competition between charge and spin fluctuations, and favor a charge-ordered state. 
Below, we discuss the results obtained using the simple bath model  (Fig.~\ref{fig:bath_simple}). 

\begin{figure}
\centering
\includegraphics[width=\columnwidth]{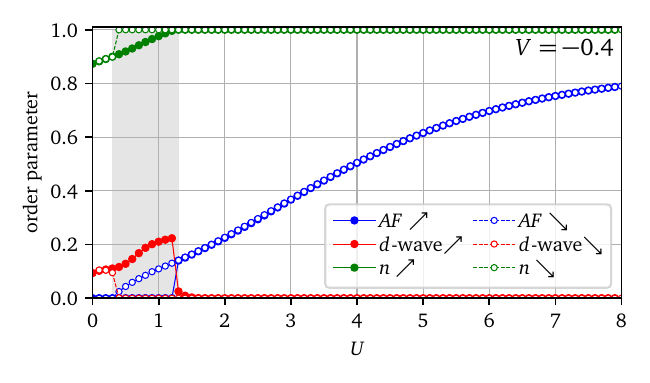}
\includegraphics[width=\columnwidth]{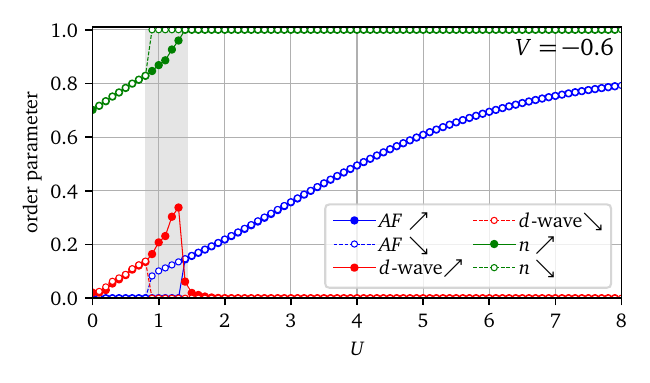}
\caption{First-order phase transition from $d$-wave superconductivity (indicated by filled/open red circles) to antiferromagnetism (AF, indicated by filled/open blue circles), as a function of the repulsive local interaction $U$, at fixed $V=-0.4$ (top) and $V=-0.6$ (bottom), and fixed chemical potential $\mu=U/2+4V$ (particle-hole symmetric point).
The simple impurity model (Fig.~\ref{fig:bath_simple}) is used.
The transition is accompanied by a deviation in the number density (indicated by filled/open green circles) from the half-filled value $n=1$, meaning that we are entering a phase separated regime. This may also explain the rapid suppression of superconductivity for smaller values of $U$ for a more negative interaction $V$. 
The dashed (solid) curves of each color depict the behavior of the different quantities for decreasing (increasing) $U$, respectively.
The prominent region of hysteresis between the two curves confirms the order of the transition.\label{fig:V<0funcofU}}
\end{figure}

\begin{figure}
\centering
\includegraphics[width=\columnwidth]{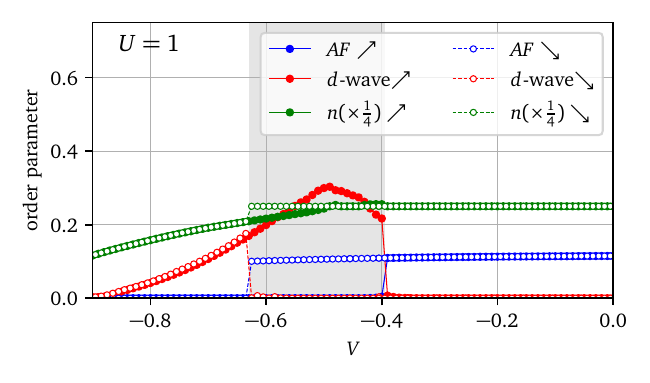}
\includegraphics[width=\columnwidth]{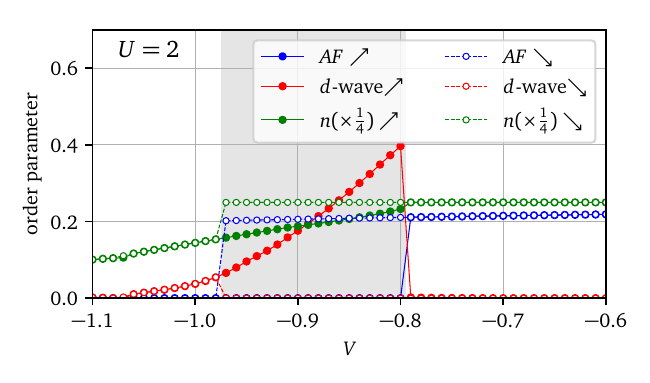}
\caption{First-order phase transition from antiferromagnetism (AF) (indicated by filled/open blue circles) to $d$-wave superconductivity (indicated by filled/open red circles), for increasingly attractive $V$, followed by a rapid suppression in the superconducting order parameter, for on-site interaction $U=1$ (top) and $U=2$ (bottom).
The simple impurity model (Fig.~\ref{fig:bath_simple}) is used.
The transition is accompanied by a deviation in the number density (indicated by filled/open green circles) from the half-filled value $n=1$.
The dashed (solid) curves of each color depict the behavior of different quantities for decreasing/more negative (increasing/less negative) $V$, and we find significant hysteresis.
For larger repulsive interactions $U$, the transition is found to occur at a critical value of $V$ that is more attractive.\label{fig:V<0funcofV}}
\end{figure}

\subsubsection{$V<0:$}

For a fixed attractive nearest-neighbor interaction $V$, as the strength of the local repulsive interaction $U$ decreases, the system undergoes a first-order phase transition from antiferromagnetism to $d$-wave superconductivity. This is accompanied by a deviation in the electron density from its half-filled limit, which can be attributed to the effects of phase separation, discussed in more detail in the next subsection.
Each of the order parameters is plotted for both increasing and decreasing $U$, and the region of hysteresis between the two curves indicates that the transition is first-order in nature.
We have verified that other pairing symmetries, such as extended $s$-wave and $p$-wave, do not compete with $d_{x^2-y^2}$ pairing in this regime.
The results of our analysis are shown in Fig.~\ref{fig:V<0funcofU}.
Likewise, 
an antiferromagnetic order is destabilized in favor of $d-$wave superconductivity for an attractive $V$, at a fixed repulsive $U\sim t$, with significant hysteresis between the curves obtained for increasing/decreasing $V$. The latter state  is then rapidly suppressed due to the effect of phase separation.
The results are shown in Fig.~\ref{fig:V<0funcofV}. 

\begin{figure}
\centering
\includegraphics[width=\columnwidth]{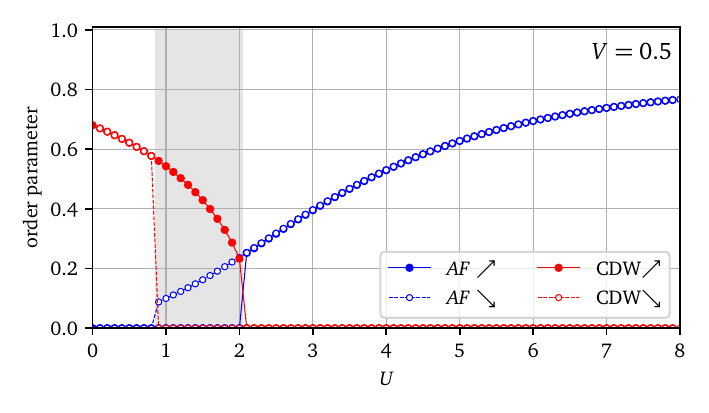}
\includegraphics[width=\columnwidth]{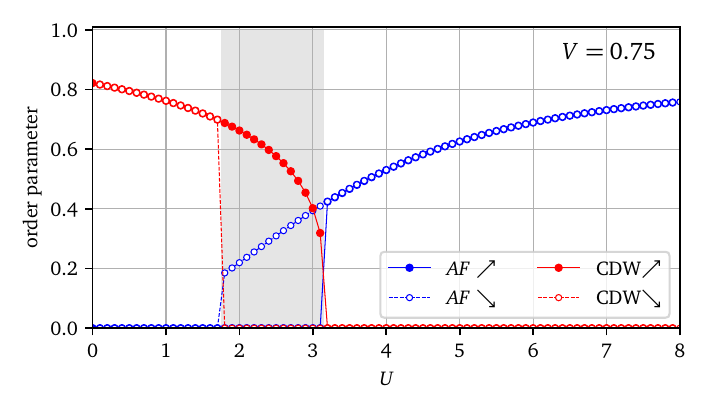}
\caption{First-order phase transition from a charge-density wave (CDW) order (indicated by filled/open red circles) to antiferromagnetism (indicated by filled/open blue circles), at half-filling, as a function of the local repulsive interaction $U$, for $V=0.5$ (top) and $V=0.75$ (bottom).
The simple impurity model (Fig.~\ref{fig:bath_simple}) is used.
The dashed (solid) curves of each color depict the behavior of the order parameters for decreasing (increasing) $U$, and exhibit significant hysteresis.
As the repulsive $V$ becomes stronger, the transition is found to occur at a larger value of $U$, the CDW order parameter increases considerably in magnitude, and the region of hysteresis is somewhat enhanced.\label{V>0funcofU}}
\end{figure}

\begin{figure}
\begin{centering}
\includegraphics[width=\columnwidth]{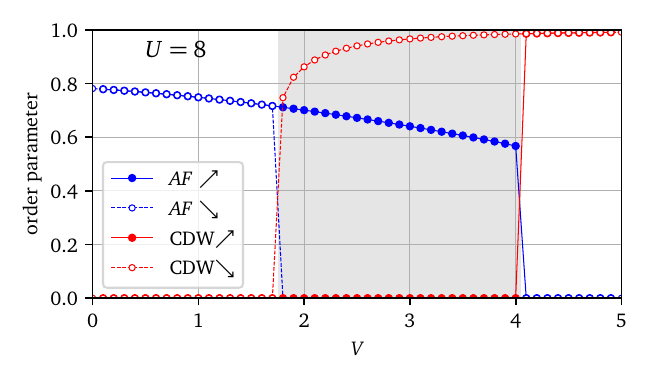}
\includegraphics[width=\columnwidth]{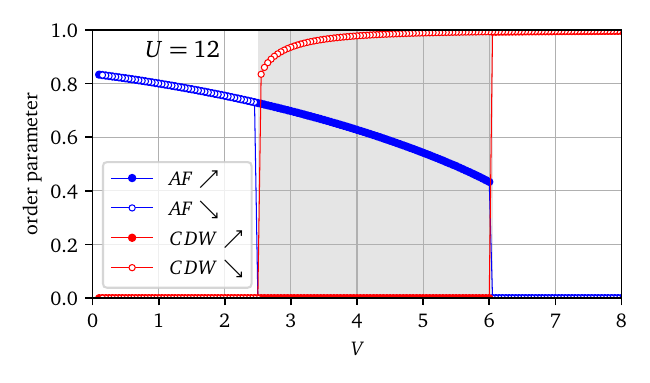}
\par\end{centering}
\caption{First-order phase transition from antiferromagnetism (indicated by filled/open blue circles) to charge-density wave (CDW) order (indicated by filled/open red circles), at half-filling, as a function of the repulsive interaction $V$ for fixed $U$, with $U=8$ (top) and $U=12$ (bottom).
The simple impurity model (Fig.~\ref{fig:bath_simple}) is used.
The dashed (solid) curves of each color depict the behavior of the order parameters for decreasing (increasing) $V$, and exhibit considerable hysteresis.
As $U$ increases, the transition occurs at a larger critical value of $V$, and the antiferromagnetic order parameter increases in magnitude.\label{V>0funcofV}}
\end{figure}

\subsubsection{$V>0:$}

For repulsive nearest-neighbor interactions $V$, we do not expect to find any superconducting orders at half-filling in the strong-coupling limit $U\gg t$, and instead focus on studying the competition between charge- and spin-density-wave orders.
At fixed $V>0$, we observe a first-order phase transition from a  charge-density wave (CDW) to an antiferromagnetic (AF) state, as a function of increasing $U$.
Likewise, for a large repulsive $U$, the system undergoes a phase transition from antiferromagnetism to CDW, as a function of the repulsive $V$.
In both cases, a large region of hysteresis is observed between the results obtained for increasing and decreasing values of the respective interaction couplings.
The results of our analysis are shown in Figs~\ref{V>0funcofU} and \ref{V>0funcofV}, respectively.

We do not present the corresponding results for the more general bath model (Fig.~\ref{fig:bath_general}) here, as they are found to be qualitatively similar to those obtained for the simple model. The key differences, that are sometimes observed, include a) an increase/decrease in the strength of the $d-$wave order parameter close to the transition, b) a smaller region of hysteresis, c) a small shift in the position of the transition, particularly as a function of $V$ for fixed $U$.
\subsection{Phase diagram as a function of density}

Next, we examine the phase diagram of the model over a continuous range of densities, for $U>0$ and attractive/repulsive $V$.
For $V>0$, we once again limit ourselves to the strong-coupling limit $U\gg t$.
For $V<0$, we focus on studying the effect of an attractive extended interaction, with a local repulsion $U$ controlling the extent of phase separation. 

\begin{figure*}
\begin{centering}
\includegraphics[width=2\columnwidth]{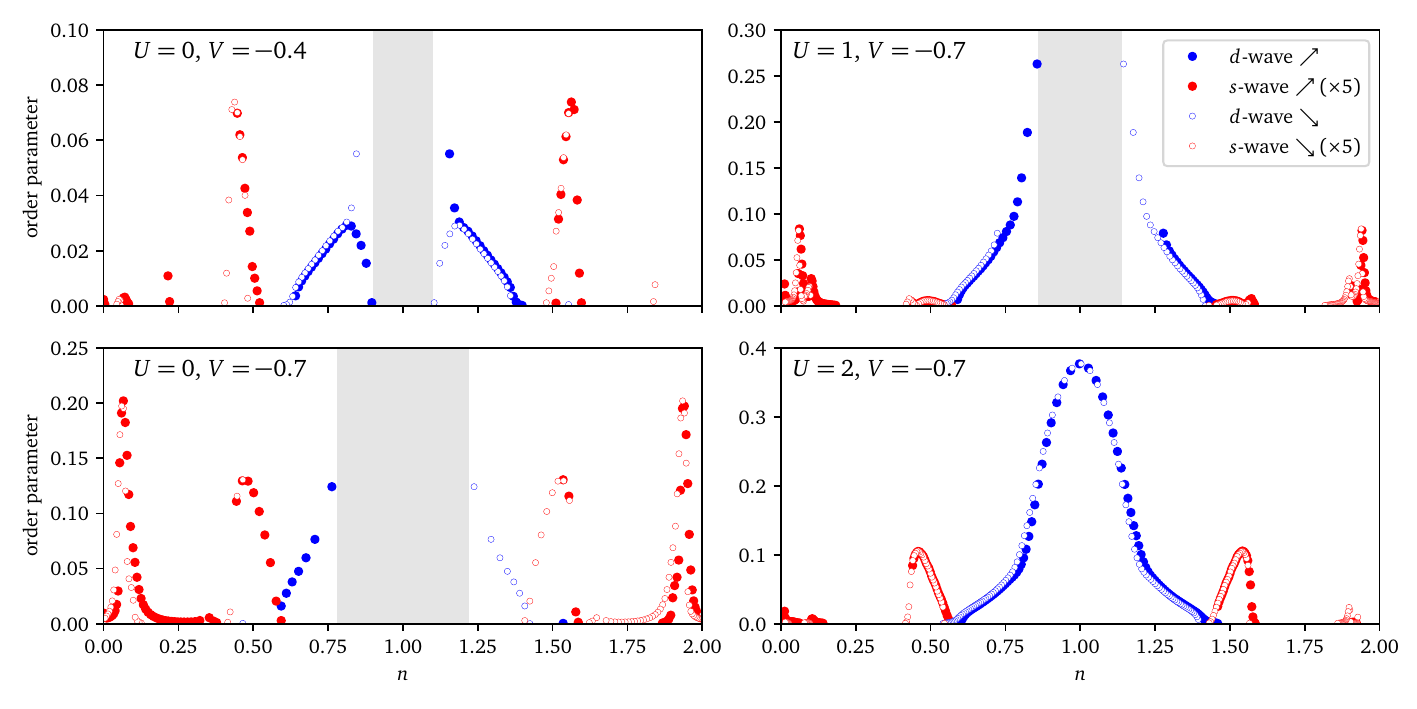}
\par\end{centering}
\caption{Superconducting order parameter of the EHM with attractive nearest-neighbor interactions, as a function of density $n$, from $n=0$ to $2$ for the simple bath model (Fig.~\ref{fig:bath_simple}).
Close to the half-filled value $n=1$, we find signatures of phase separation, indicated by a gap in the curve 
over a range of densities, caused by a jump in the compressibility $\partial n/\partial\mu$ (as shown in Fig.~\ref{mun}).
For smaller (larger) fillings, symmetrical and sharply defined regions of $d$-wave superconductivity (represented by filled/open blue circles) are followed by disconnected patches of extended $s$-wave order (represented by filled/open red circles), which appear only beyond a critical attractive value of $V$.
\label{V<0funcoffilling}}
\end{figure*}

\begin{figure*}
\begin{centering}
\includegraphics[width=2\columnwidth]{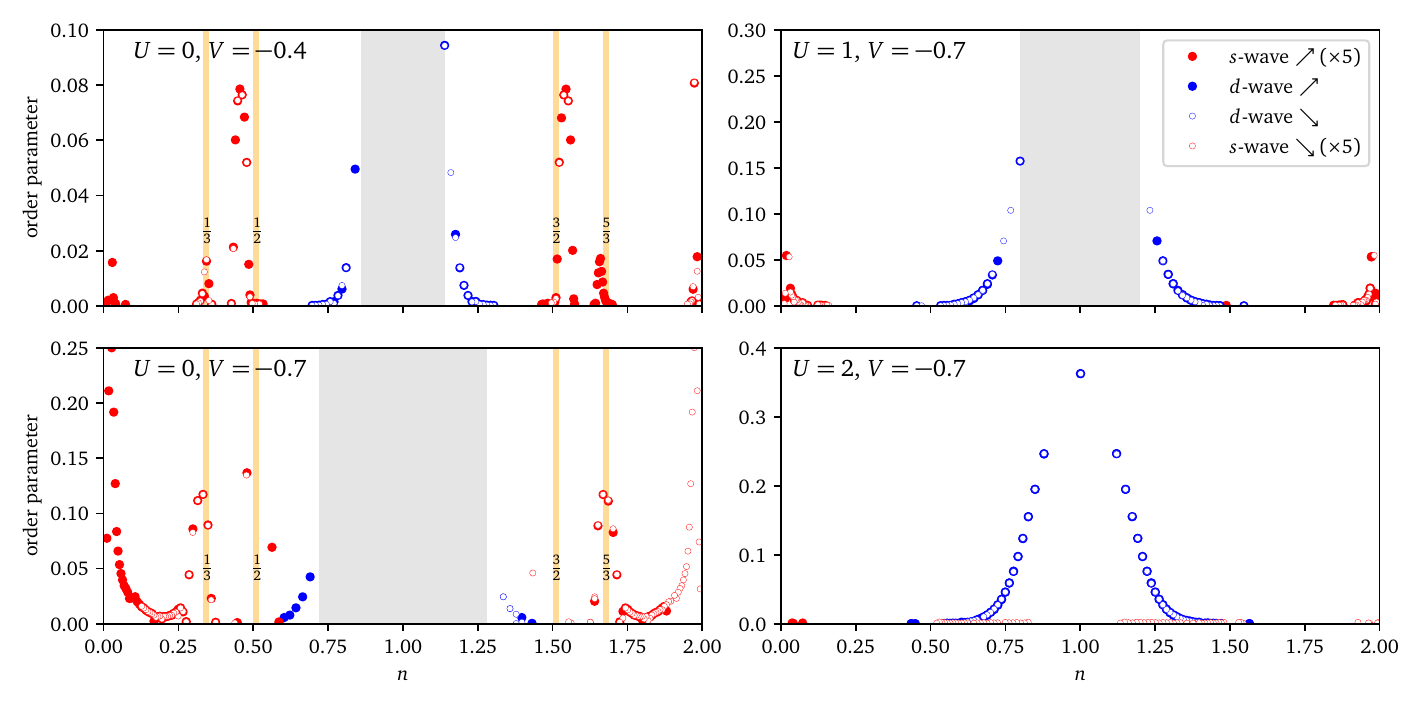}
\par\end{centering}
\caption{Superconducting order parameter of the EHM with attractive nearest-neighbor interactions, as a function of density $n$, from $n=0$ to $2$ for the general bath model (Fig.~\ref{fig:bath_general}). The overall behavior of the $d-$ and extended $s$-wave patches are similar to the corresponding result for the simple bath model. However, note that the structure of the $s$-wave order parameter has changed, with a more extended region near quarter-filling, and an additional patch near $1/3-$filling. For $U=0,V=-0.7$, the phase separation region extends all the way to quarter-filling, and the corresponding superconducting patches are almost absent, and asymmetric about $n=1$. Moreover, the new $s$-wave order parameter becomes unambiguously weaker as the repulsive $U$ increases, and is completely absent for $U=1$ and $U=2$, thus resolving the question of the nonmonotonous behavior of the $s$-wave order parameter in the simple bath model. 
\label{V<0funcoffilling1}}
\end{figure*}

\begin{figure}
\begin{centering}
\includegraphics[width=0.9\columnwidth]{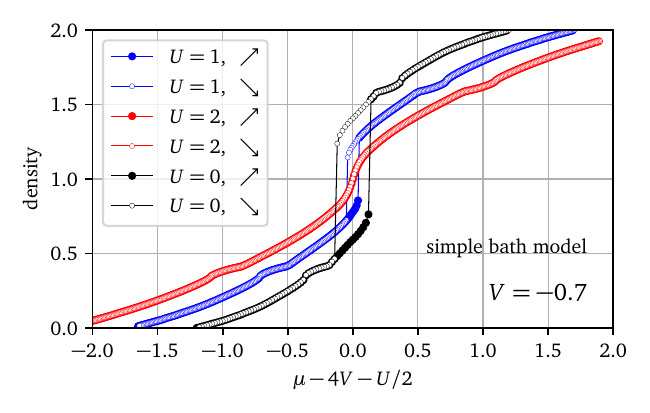}
\par\end{centering}
\caption{Number density $n$ as a function of the chemical potential $\mu$ (measured with respect to its particle-hole invariant value, $\mu_{c}=U/2+4V$) for an EHM with attractive nearest-neighbor interactions, over a range of values of $U\protect\geq0$ and $V<0$
for the simple bath model (Fig.~\ref{fig:bath_simple}).
On either side of half-filling ($\mu=\mu_{c})$, we find symmetrical jumps in the compressibility $\partial n/\partial\mu$ enclosing a region of hysteresis, which corresponds to the coexistence of two different uniform-density solutions.
This is interpreted as the region of phase separation.
The red, blue and black filled/open circles represent the behavior for various values of $U$ for $V=-0.7$, and demonstrate that while a sufficiently attractive interaction $V$ favors phase separation,
a stronger on-site repulsion $U$ is detrimental to it.\label{mun}}
\end{figure}

\begin{figure}
\begin{centering}
\includegraphics[width=0.9\columnwidth]{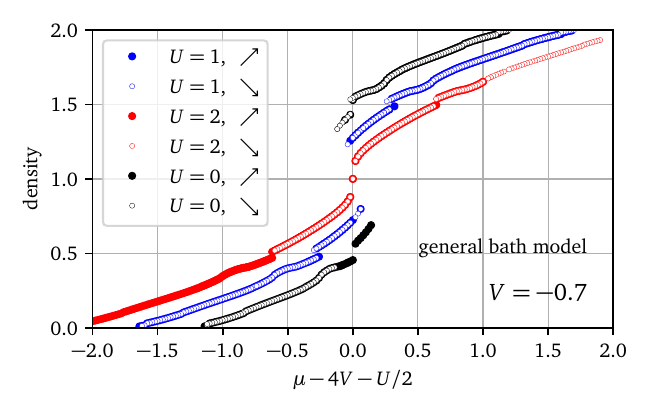}
\par\end{centering}
\caption{Number density $n$ as a function of the chemical potential $\mu$ (measured with respect to its particle-hole invariant value, $\mu_{c}=U/2+4V$) for the EHM with attractive nearest-neighbor interactions, over a range of values of $U\protect\geq0$ and $V<0$
for the general bath model (Fig.~\ref{fig:bath_general}). The behavior is very similar to that observed in the simple bath model, with the most notable difference being the appearance of symmetric jumps in the number density $n$, close to quarter-filling, for each of the curves. 
\label{mun1}}
\end{figure}

\subsubsection{$V<0:$}

Let us now discuss the different phases that are supported by the model as a function of density. Close to half-filling, we find a region of phase separation, indicated by a jump in the density, flanked by symmetrical pockets of $d_{x^2-y^2}$ pairing, which decay rapidly as a function of density. 
For further smaller (larger) fillings, an extended $s$-wave order appears in the form of disconnected patches, near quarter-filling and at very small (large) densities.
Interestingly, the variation of the extended $s-$wave order parameter as a function of $U$ and $V$ are found to be different for the simple bath model and the more general one. In the case of the simple model (see Fig.~\ref{V<0funcoffilling}), we find small regions of extended $s-$wave superconductivity near quarter-filling, 
that vary non-monotonously as a function of $U$. Only for sufficiently attractive $V$, nearly symmetrical patches of extended $s$-wave order also appear close to the band edges.
The corresponding results for the general bath model are illustrated in Fig.~\ref{V<0funcoffilling1}. While the overall magnitude of the $s$-wave order parameter turns out to be smaller than in the previous case,  its shape is more extended at quarter-filling, with two patches appearing next to each other, which, interestingly, appear close to fillings of $1/3$ and $1/2$, respectively. While it is tempting to blame the $n=1/2$ feature on a commensurate finite-size effect on a 4-site cluster, this is less obvious for the $n=1/3$ feature. The superconductivity also clearly becomes stronger as a function of $V<0$. 
Notably, the $s-$wave order is clearly absent for both $U=1$ and $U=2$, thus eliminating the confusion caused by the aforementioned non-monotonous variation in the case of the simple model. 

To better characterize the region of phase separation, we examine the behavior of the number density $n$ as a function of the chemical potential $\mu$, measured with respect to its particle-hole symmetric value $\mu_{c}=U/2+4V$.
On either side of $\mu=\mu_{c}$, we find symmetrical jumps in the compressibility $\partial n/\partial\mu$, enclosing a region of hysteresis in the $\mu-n$ curve, depicted in Fig.~\ref{mun}, where two uniform-density solutions coexist. 
Within our approach, this is interpreted as the region of phase separation, and is found to shrink under the influence of stronger local repulsive interactions $U$, and expand when $V$ becomes more attractive.
The corresponding results for the general bath model are depicted in Fig.~\ref{mun1}. The two sets of results are qualitatively similar, except for symmetric jumps observed in the number density $n$ near quarter-filling in the latter case. 
We note that the jumps occur only for the model with the larger number of bath parameters, and are the most prominent for $U=0,V=-0.7$, where the phase separation region extends all the way to quarter-filling, becoming progressively smaller for $U=1$ and $2$. It is plausible that phase separation might lead to the appearance of multiple jumps in the density, at half-filling as well as quarter-filling. Moreover, a finite-size effect would have been even more obvious in the simple bath model, where these jumps are found to be absent. The origin of the jumps is currently unclear to us. 

The appearance of a phase separated state for sufficiently attractive interactions is a familiar result~\citep{micnas_superconductivity_1989,dagotto_superconductivity_1994,su_phase_2004,van_loon_extended_2018,yao_determinant_2022,chen_superconducting_2022}, which has received attention from other groups, including very recently~\citep{sousa-junior_phase_2023}, but the characterization of the region of phase separation tends to depend on the method used for the analysis, and whether it is capable of handling a non-uniform distribution of particles.

\begin{figure}
\begin{centering}
\includegraphics[width=0.9\columnwidth]{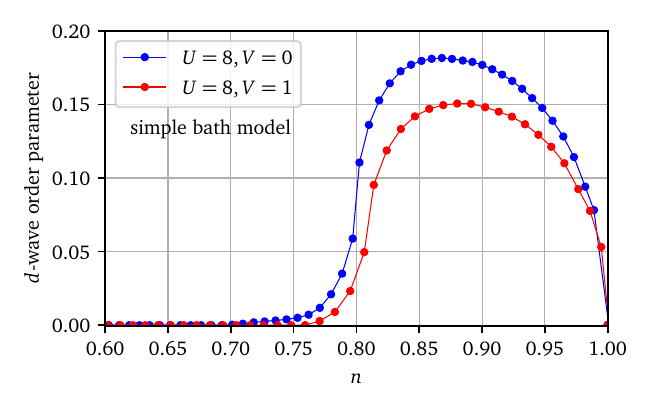}
\par\end{centering}
\caption{Superconducting $d-$wave order parameter of the EHM with repulsive nearest-neighbor interactions in the strong-coupling limit, i.e., at $U=8t$.
The Mott insulating state at half-filling is destabilized in favor of $d_{x^2-y^2}$ pairing, upon hole doping.
The dome-like region of $d$-wave superconducting order is observed for $V=0$ (indicated by the solid blue curve) and is somewhat suppressed for non-zero repulsive $V$ (indicated by the solid red curve).
No other superconducting orders are found to be stabilized in this region.\label{V>0funcoffilling}}
\end{figure}
\begin{figure}
\begin{centering}
\includegraphics[width=0.9\columnwidth]{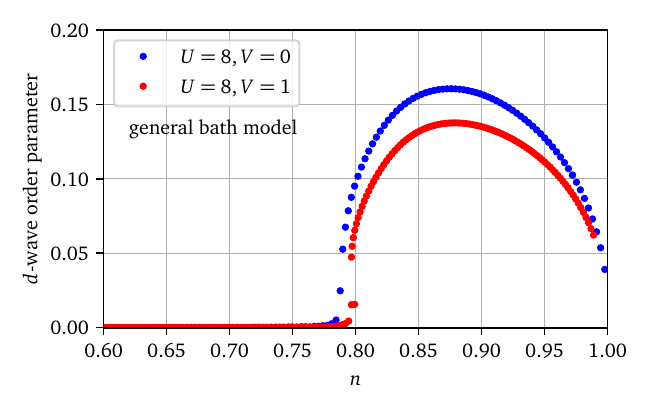}
\par\end{centering}
\caption{Superconducting $d-$wave order parameter of the EHM with repulsive nearest-neighbor interactions at intermediate coupling ($U=8t$) using the general bath model (Fig.~\ref{fig:bath_general}).
The behavior is qualitatively similar to that obtained in the simple model, with a slight difference in the magnitudes of the $d$-wave order parameter. The most noticeable difference between the two bath models is the relatively sharp transition into and out of the $d-$ wave superconducting phase. \label{V>0funcoffilling1}}
\end{figure}

\subsubsection{$V>0:$}

At half-filling, for $U=8t$, the large on-site interaction freezes the charge degree of freedom, and the ground state is a Mott insulator.
Hole doping is found to destabilize the magnetic order, and drive the system towards a $d$-wave superconducting phase.
We encounter a dome-shaped region of $d$-wave superconductivity for $V=0$, which is suppressed at smaller densities, where no competing superconducting orders are found to be stabilized in our analysis.
Upon introducing a repulsive $V\sim t$, the superconducting order remains stable, but is somewhat suppressed.
The results are depicted in Fig.~\ref{V>0funcoffilling}.
The corresponding results for the general bath model are depicted in Fig.~\ref{V>0funcoffilling1}. The two sets of results are qualitatively similar, with the most noticeable difference being the relatively sharper transition to and from the $d$-wave ordered state in the latter case.
These results are consistent with the picture of superconductivity mediated by short-range spin fluctuations in a doped Mott insulator~\cite{kyungPairing2009, senechalResilience2013, kowalskiOxygen2021}.

\section{Discussion and conclusions}

\begin{figure}
\begin{centering}
\includegraphics[width=0.95\columnwidth]{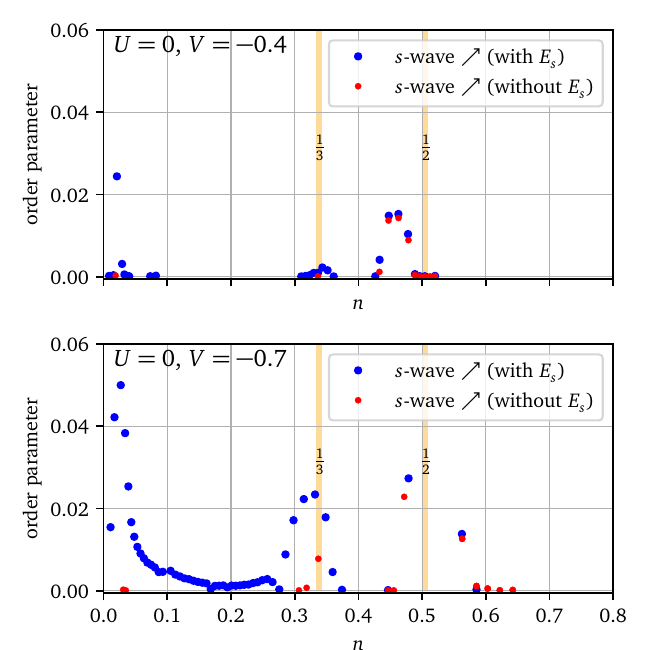}
\par\end{centering}
\caption{The figure shows the behavior of the extended $s-$wave order parameter as a function of the number density $n$, with and without the inclusion of the self-consistent anomalous mean-field parameter $E_{s}$ (see Appendix), for $\rm{U=0},\rm{V=-0.4}$ (above) and $\rm{U=0},\rm{V=-0.7}$ (below). Clearly, some of the regions with a nontrivial $s-$wave order parameter are found to be absent when $E_{s}$ is not included. Upon considering a stronger attractive $V$, these regions tend to reappear, but are suppressed in magnitude in the absence of $E_{s}$.
\label{U0V-0.4_withandwithoutSm}}
\end{figure}

In summary, we have studied the phase diagram of the extended Hubbard model, for both attractive and repulsive nearest-neighbor interactions, using a combination of Cluster Dynamical Mean Field Theory (CDMFT), with a dynamical Hartree-Fock approximation for treating inter-cluster interactions.
We examine possible phase transitions at half-filling, as well as the dominant phases that are stabilized as a function of density.
At the particle-hole invariant chemical potential, which corresponds to a half-filled band in the absence of phase separation, the antiferromagnetically ordered state undergoes a first-order phase transition to $d$-wave superconductivity for a critical attractive interaction $V$.
Stronger attractive extended interactions also tend to induce phase separation, which manifests itself in the form of a gradual deviation of the density from its half-filled limit, for a fixed chemical potential.
For a sufficiently strong repulsive interaction $V$, a charge-density wave order is stabilized at half-filling.
As a function of density, a phase separated state near the half-filled point is flanked by symmetrical regions of $d$-wave superconductivity, that decay sharply as a function of density, and disconnected patches of extended $s$-wave order at smaller (larger) band fillings.
For the case of repulsive non-local interactions, in the strongly coupled limit, the Mott insulator at half-filling gives way to a dome-shaped region of $d$-wave superconductivity, upon hole doping, which is expected on physical grounds.
No other competing superconducting orders are found to be stabilized in this region of parameter space.

For the most part, our results are found to be qualitatively consistent with the existing literature.
The transition between antiferromagnetism and CDW at half-filling, for repulsive interactions, has been predicted by several previous studies~\citep{aichhorn_charge_2004,chattopadhyay_phase_1997,murakami_possible_2000,philoxene_spin_2022,pudleiner_competition_2019,rosciszewski_2003,sousa-junior_phase_2023,terletska_charge_2017,terletska_charge_2018,terletska_dynamical_2021,yao_determinant_2022}, although the critical interaction strength typically depends on the method of analysis.
For densities away from half-filling, there have also been some predictions of $d_{xy}$ pairing, that appears beyond the region of $d_{x^2-y^2}$ pairing, for repulsive extended interactions 
~\citep{onari_phase_2004,huang_unconventional_2013}.
We do not find such a state in our analysis.
The phase diagrams obtained from self-consistent mean-field theory based analyses tend to prominently feature $d$-wave superconductivity at half-filling, with a continuous region of extended $s$-wave order at smaller densities, along with a region of coexistence between the two, i.e., $s+id$ pairing~\citep{micnas_extended_1988,micnas_superconductivity_1988}.
In our analysis, we do not usually see a coexistence between $d$- and extended $s$-wave orders. In the simple model, such a coexistence is observed only in those regimes where both the interactions $U>0$ and $V<0$ are sufficiently strong, and comparable in magnitude.
This may be due to the fact that the superconducting orders found in our analysis are fairly weak, and the significant attractive interactions that are, therefore, needed for stabilizing overlapping regions of $d$- and extended $s$-wave orders, would also lead to a larger region of phase separation.
This effect can only be compensated by including a sufficiently large repulsive local interaction. On the other hand, we have not been able to verify a similar coexistence of the orders for the general bath model, due to the rapid suppression of the extended $s-$wave order, near quarter-filling, with an increase in $U$. 
Some studies have also suggested the possibility of $p$-wave superconductivity, especially at half-filling~\citep{chen_superconducting_2022}, and for intermediate hole doping, beyond the region of $d$-wave superconducting order~\citep{micnas_extended_1988,micnas_superconductivity_1988,huang_unconventional_2013}.
We do not find signatures of $p$-wave superconductivity in the parameter regimes that we study.
Some of our results at half-filling are found to be qualitatively consistent with a recent study on the extended Hubbard model using the determinantal Quantum Monte Carlo technique~\citep{sousa-junior_phase_2023}, which also reports the transitions between $d$-wave superconductivity and AFM, as well as between phase separation and $d$-wave, that we observe in our analysis.
In addition, the authors of the aforementioned paper also explore other quadrants of the $U-V$ phase diagram, including the case where $U<0$, which we do not take into account, since the repulsive component of the Coulomb interaction is always expected to be present in a realistic situation.

In contrast to ordinary mean-field theory, our approach takes the intra-cluster fluctuations into account exactly, and is therefore expected to give more reliable quantitative results. 
In particular, ordered phases are weaker in this approach than in ordinary mean-field theory.
At the same time, it should be noted that we only take into account spatial fluctuations within small clusters, and the accuracy of the method is controlled by the size of the clusters used.
To illustrate the importance of including the effect of the inter-cluster interactions self-consistently, which are usually disregarded in cluster-based approaches, we have compared the behavior of the superconducting $d-$ and extended $s-$wave orders as a function of density $n$, for an attractive $V$ (see  Fig.~\ref{U0V-0.4_withandwithoutSm}) in the presence and absence of the anomalous mean-field parameters (which we refer to as $E_{d}$ and $E_{s}$ respectively). Certain regions of the extended $s-$wave order, that we observe in our analysis, disappear entirely in the absence of the self-consistent anomalous mean field  parameter $E_{s}$. These regions tend to reappear, but with a smaller amplitude, when the attractive $V$ is sufficiently strong. Likewise, in the case of $d-$wave superconductivity, we find that the superconducting order parameter is negligible when $E_{d}$ is absent, and tends to reappear, with a much smaller amplitude, when the repulsive $U$ is increased. 
Our approach is more suitable for making predictions about the thermodynamic limit than exact diagonalization studies on finite-sized clusters, since only the self-energy is limited by the cluster size.
Some recent studies have explored the possibility of magnetic states characterized by ordering wave vectors that are incommensurate with the lattice periodicity~\citep{scholle_comprehensive_2023} in the two-dimensional Hubbard model, for electron densities below half-filling, where the antiferromagnetic state becomes unstable.
Our approach is unsuitable for identifying such incommensurate charge and spin orders.
Our method does not suffer from fundamental restrictions on its applicability in any particular parameter regime, and allows us to study the behavior of the model as a continuous function of doping, rather than by focusing on specific densities, as has been done in many previous studies.
In the future, this method could be  potentially useful for analysing more complicated models, including those with spin-orbit interactions.

\acknowledgements

S.K. acknowledges financial support from the Postdoctoral Fellowship from Institut Quantique, and from UF Project No. P0224175 - Dirac postdoc fellowship, sponsored by the Florida State University National High Magnetic Field Laboratory (NHMFL). D.S. acknowledges support by the Natural Sciences and Engineering Research Council of Canada (NSERC) under grant RGPIN-2020-05060. Computational resources were provided by the Digital Alliance of Canada and Calcul Québec.

\appendix*

\begin{figure}
\begin{centering}
\includegraphics[width=0.9\columnwidth]{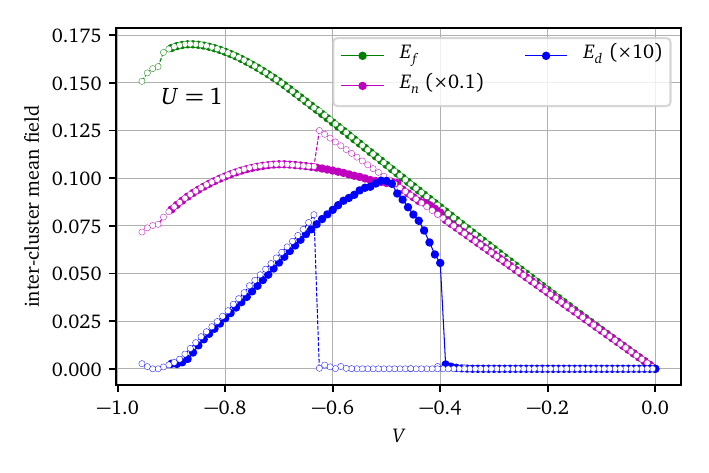}
\par\end{centering}
\caption{Inter-cluster Hartree-Fock mean fields for the solutions shown in the top panel of Fig.~\ref{fig:V<0funcofV}.
$E_d$ is the eigen-operator associated with $d$-wave superconductivity, $E_f$ with the nearest-neighbor
kinetic operator $f_{\rv\rv'\sigma\sigma}$ and $E_n$ with the density $n$ (basically a shift in the chemical potential induced by $V$). The mean-field $E_s$  associated with extended $s$-wave superconductivity is negligible over almost the entire range of $V$, since this is at half-filling, except at significantly attractive $V$ (due to phase separation).
Note the very different scales (the superconducting mean field is much magnified). The filled and empty circles denote the  results for increasing (less negative) and decreasing (more negative) $V$, respectively. 
\label{U1_funcofVneg_halffilled_MF}}
\end{figure}

\section{}

The extended interaction term can be rewritten as 
\begin{multline*}
 \frac{1}{2}\sum_{\rv,\rv',\sigma,\sigma'}V_{\rv\rv'}n_{\rv\sigma}n_{\rv'\sigma'}\\
 =\frac{1}{2}\sum_{\rv,\rv',\sigma,\sigma'}V_{\rv\rv'}^{\rm{c}}n_{\rv\sigma}n_{\rv'\sigma'}+\frac{1}{2}\sum_{\rv,\rv',\sigma,\sigma'}V_{\rv\rv'}^{\rm{ic}}n_{\rv\sigma}n_{\rv'\sigma'}
\end{multline*}
where $\rv,\rv'$ refer to the lattice sites, and $n_{\rv\sigma}$ is the number of particles at site $\rv$ with spin $\sigma$.
Here $V_{\rv\rv'}^{\rm{c}}$ and $V_{\rv\rv'}^{\rm{ic}}$ refer to the intra-cluster and inter-cluster parts of the interaction.
Inspired by Wick's theorem, we decompose the inter-cluster part of the interaction into Hartree, Fock and anomalous channels, as follows:
\begin{multline}\label{eq:MF}
\frac{1}{2}\sum_{\rv,\rv',\sigma,\sigma'}V_{\rv\rv'}^{\rm{ic}}n_{\rv\sigma}n_{\rv'\sigma'}
=\kern-1em\sum_{\rv,\rv',\sigma,\sigma'}V_{\rv\rv'}^{\rm{ic}}(n_{\rv\sigma}\bar n_{\rv'\sigma'}-\frac{1}{2}\bar n_{\rv\sigma}\bar n_{\rv'\sigma'})\\
-\sum_{\rv,\rv',\sigma,\sigma'}V_{\rv\rv'}^{\rm{ic}}(f_{\rv\rv'\sigma\sigma'}\fb_{\rv\rv'\sigma\sigma'}^{*}-\frac{1}{2}\fb_{\rv\rv'\sigma\sigma'}^{*}\fb_{\rv\rv'\sigma\sigma'})\\
+\frac{1}{2}\sum_{\rv,\rv',\sigma,\sigma'}V_{\rv\rv'}^{\rm{ic}}(\Delta_{\rv\rv'\sigma\sigma'}\Deltab_{\rv\rv'\sigma\sigma'}^{*}+\Delta_{\rv\rv'\sigma\sigma'}^{\dagger}\Deltab_{\rv\rv'\sigma\sigma'}\\
-\Deltab_{\rv\rv'\sigma\sigma'}\Deltab_{\rv\rv'\sigma\sigma'}^{*})
\end{multline}
where the operators are defined as $n_{\rv\sigma}\equiv c_{\rv\sigma}^{\dagger}c_{\rv\sigma},f_{\rv\rv'\sigma\sigma'}\equiv c_{\rv\sigma}^{\dagger}c_{\rv'\sigma'}$ and $\Delta_{\rv\rv'\sigma\sigma'}\equiv c_{\rv\sigma}c_{\rv'\sigma'}$.
Note that the applicability of Wick's theorem is not exact in this case, as we are considering a model which already includes on-site interactions, but must be considered as an {\it{ad hoc}} Ansatz.
In other words, at a fundamental level, we are not assuming that the ground state of the system is a Slater determinant.
We are rather resting on a variational principle for the self-energy~\cite{potthoff2003} on which CDMFT is formally based.

The sum over sites $\rv,\rv'$ is taken over the whole lattice.
But the average $\bar n_{\rv\sigma}$ will be assumed to have the periodicity of the cluster, i.e., 
$\bar n_{\rv+\Rv\sigma} = \bar n_{\rv\sigma}$ where $\Rv$ belongs to the super-lattice.
In addition, the two-site averages $\fb_{\rv\rv'\sigma\sigma'}$ and $\Deltab_{\rv\rv'\sigma\sigma'}$ are assumed to depend only on the relative position $\rv-\rv'$.
The mean-field inter-cluster interaction \eqref{eq:MF} is then a one-body contribution to the Hamiltonian with the periodicity of the super-lattice, and contains both intra-cluster and inter-cluster terms, whereas the purely intra-cluster part $V_{\rv\rv'}^{\rm{c}}$ retains its fully correlated character.

For a four-site cluster, we have a total of eight bonds between neighboring clusters, along the $x$ and $y$ directions, with two spin combinations ($\sigma,\sigma'$) per bond, where we consider spin-parallel combinations for the Fock terms (in the absence of spin-dependent hopping) and spin-antiparallel combinations for the anomalous terms. 
In practice, we only consider physically relevant combinations of operators defined on different sites/bonds for our analysis (such as those compatible with a $d-$wave or an extended $s-$wave order).
As an illustration of this, let us consider the pairing fields $\Delta$ defined on all of these bonds, which we denote by the labels $i=1-16$ (including different bond and spin combinations). 

The mean-field Hamiltonian can be written as 
\begin{equation}
\frac{V}{2}\sum_{i,j}(\Deltab_{i}^{*}M_{ij}\Delta_{j}+\Delta_{i}^{\dagger}M_{ij}\Deltab_{j}-\Deltab_{i}^{*}M_{ij}\Deltab_{j})\label{eq:1}
\end{equation}
where $i,j=(\rv,\rv',\sigma,\sigma')$ and the matrix $M_{ij}$ describes the combinations of the pairing fields defined on different bonds which appear in the Hartree-Fock decomposition of the intercluster interactions. The matrix $M$ turns out to be an identity matrix for the Fock and pairing fields $f$ and $\Delta$ respectively, but the corresponding matrix for the Hartree fields $n$ is off-diagonal.

Defining the eigen-combinations of the pairing fields by 
\begin{equation}
E_{\alpha}=U_{\alpha i}\Delta_{i}
\end{equation}
and the eigenvalues of the matrix $M$ by $\lambda_{\alpha}$, such that
\begin{equation*}
M_{ij}=\sum_{\alpha,\beta}U_{\alpha i}^*\lambda_{\alpha}\delta_{\alpha\beta}U_{\beta j}
\end{equation*}
we can rewrite Eq.~\eqref{eq:1}, above, as 
\begin{equation}
\frac{V}{2}\sum_{\alpha}\lambda_{\alpha}(\bar E_{\alpha}^{*}E_{\alpha}+E_{\alpha}^{\dagger}\bar E_{\alpha}-\bar E_{\alpha}^{*}\bar E_{\alpha}) \end{equation}
The mean-field values $\bar E_{\alpha}$ of the relevant eigen-combinations $E_{\alpha}$ of the pairing operators defined on different nearest-neighbor bonds are obtained self-consistently within the CDMFT loop, and likewise for the other mean fields that are the appropriate eigen-combinations of $\bar n_{\rv\sigma}$ and $\fb_{\rv\rv'\sigma\sigma'}$.

\bibliographystyle{apsrev4-1}
%


\end{document}